\documentclass[article,aps,amsmath,nofootinbib,english,11pt]{revtex4}
\usepackage{graphics,color,array,dcolumn}
\usepackage{graphicx}
\usepackage{amssymb}
\usepackage{xspace}
\usepackage{color}

\def\be{\begin{equation}}
\def\ee{\end{equation}}
\def\ba{\begin{eqnarray}}
\def\ea{\end{eqnarray}}

\makeatother
\begin{document}

\title{Asymptotic Safety, Emergence and Minimal Length}
\bigskip
\author{Roberto Percacci}
\email{percacci@sissa.it}
\affiliation{
SISSA, via Bonomea 265, I-34136 Trieste
\\
and 
INFN, Sezione di Trieste, Italy
}
\author{Gian Paolo Vacca}
\email{vacca@bo.infn.it}
\affiliation{INFN, Sezione di Bologna,
via Irnerio 46, I-40126 Bologna}
\pacs{}

\bigskip

\begin{abstract}
There seems to be a common prejudice that asymptotic safety is
either incompatible with, or at best unrelated to, the other topics in the title. 
This is not the case. In fact, we show that 
1) the existence of a fixed point with suitable properties
is a promising way of deriving emergent properties of gravity, and
2) {there is a sense} in which asymptotic safety implies a minimal length.
In so doing we also discuss possible signatures of asymptotic safety
in scattering experiments.
\end{abstract} 

\maketitle

\section{Introduction}

A Quantum Field Theory (QFT) is said to be an ``effective field theory'' (EFT)
if it breaks down at some energy scale, 
and ``fundamental'' or ``UV complete'' if it makes sense up to arbitrarily high energy scales.
QCD is our best example of the latter behavior.
Other pieces of the standard model, in particular the Higgs sector,
and Einstein's theory of gravity, are usually regarded as EFT's, 
because they are expected to break down near their characteristic energy scale.

Asymptotic safety is a kind of behavior that would ensure UV completeness \cite{reviews}.
Assuming that a ``theory space'' has been defined by giving a set of fields
and symmetries, and that it is parameterized by all the couplings 
\footnote{here by ``couplings'' we mean the dimensionless numbers formed
by multiplying every coupling by a suitable power of the cutoff,
see eq. (\ref{seff}) below. Also, it is sufficient to consider only the so-called
essential couplings, see \cite{reviews}.}
that may appear in the effective action, consider the RG flow in this theory space. 
Each theory corresponds to an RG trajectory.
Asymptotic safety requires that there exists a Fixed Point (FP) where
all beta functions vanish, and that the tangent space at this
point is spanned by only finitely many relevant (UV attractive)
directions, and infinitely many irrelevant (UV repulsive) directions.
One calls ``UV critical surface'' the basin of attraction of the FP
for increasing energy. 
Assuming that it is smooth in some sense, its dimension $d$ is equal
to the number of relevant couplings.
Likewise one defines the ``IR critical surface'' to be the basin of
attraction of the FP for decreasing energy.
Its dimension is equal to the number of irrelevant couplings and is generally infinite.
A theory is said to be asymptotically safe if the corresponding
trajectory lies in the UV critical surface.
Then, $d$ measurements performed
at a given energy are enough to pin down the theory, and the remaining
(infinitely many) coordinates of the trajectory are a prediction of the theory 
which can be tested against further experiments.

QCD is a special type of asymptotically safe theory, namely one
that is governed by the Gaussian (free) FP in the UV.
We know that in the Higgs model and in gravity, if we start
in the domain where perturbation theory is applicable,
namely near the Gaussian FP, we are driven away from it.
So, in these theories the perturbative expansion breaks down,
and possibly this signals a breakdown of the QFT as a whole.
Could these theories be governed by nontrivial FP's in the UV?
This is a very attractive proposition, and it has generated
a considerable amount of research.
How can one establish whether the standard model 
and/or gravity conform to this behavior?
The best way to approach this problem is from the bottom up.
One begins within the framework of EFT, with an explicit UV cutoff $k$. 
This theory, treated at tree level,
is supposed to give a good description of physical phenomena
characterized by momentum scale $k$.
Then one studies the
behavior of the (renormalized) couplings as $k$ is allowed to grow
\footnote{for this it is important to have a way of computing
beta functions which does not rely on reading the coefficients of
divergent terms in the effective action. The Wilsonian RG
provides finite answers for the beta functions of any theory without
having to discuss the UV limit first.}.
In general, if one started from a finite number of couplings,
new couplings would be generated by the flow, so in order to be sure
that one is not missing any information, one should in principle
work in the full theory space, {\it i.e.}
compute the RG of all couplings.
If some coupling blows up at a scale $k=\Lambda_{UV}$,
then that scale defines an upper bound on the range of validity
of the theory, which has therefore to be interpreted as an EFT.
If, on the other hand, one can find a trajectory for which
all couplings have a finite limit when $k\to\infty$, 
the theory is UV complete. 
Notice again that UV completeness is a property of certain trajectories: 
if a FP with the right properties exist,
there are trajectories that end at the FP in the UV and so describe
UV complete theories, and other
trajectories that do not end at the FP in the UV
and therefore describe EFT's.

Since the UV problem is more acute in gravity, this is where
most of the recent work has been focused.
Some progress in this direction has come from numerical simulations
in discretized models, as for example Regge calculus \cite{hamber}
or Causal Dynamical Triangulations (CDT) \cite{ambjorn}.
Here we shall focus mainly on the continuum approach
which is based on the application of functional RG methods \cite{frge}.
These tools were applied first to Einstein theory
\cite{reuter1,dou,souma,lauscher,litim,saueressig,eichhorn},
followed by extensions to four-derivative gravity
\cite{codello1,bms,niedermaier}
and to $f(R)$ gravity \cite{reuter2,cpr,ms}.
Specific issues have been addressed in the conformal truncation,
where one freezes the spin two degrees of freedom
\cite{creh,machado}, 
or in bimetric truncations \cite{manrique}.
In view of possible applications to the standard model,
there have been preliminary studies of Yukawa systems
\cite{giessm} and nonlinear sigma models \cite{nlsm}.
Mixed systems with gravity and minimally coupled
and/or selfinteracting matter have been
considered in \cite{griguolo,perini,largen,zzvp,narain,daum}.

Without entering into a detailed discussion of the evidence for a FP,
in the following it will be useful to have an approximate
but explicit understanding of the running of Newton's constant.
The coefficient of the Hilbert action is the square of Planck's mass:
$M_{pl}^2=\frac{1}{16\pi G}$
\footnote{we assume that special relativity and quantum mechanics
are correct in their respective domains, and we choose units
such that $c=1$ and $\hbar=1$. Then everything has dimension of
a power of mass. When we talk of a choice of units we always
mean the choice of a unit of mass.}.
In the quantum theory it is expected to diverge quadratically,
leading to a beta function of the form
\begin{equation}
\label{planckrg}
k\frac{d}{dk}M_{pl}^2=2 a k^2\ ,
\end{equation}
where $a$ is some constant. This expectation is borne out by a number of 
independent calculations, showing that the leading term in the beta function
has this behavior, with $a>0$ \cite{souma,lauscher,litim,saueressig,cpr}.
Let $\tilde G=G k^2$ be the dimensionless number 
expressing Newton's constant in units of the cutoff $k$.
Then, the beta function of $\tilde G$ has the form
\begin{equation}
k\frac{d\tilde G}{dk}=2\tilde G-32\pi a\, \tilde G^2\ .
\end{equation}
This beta function has an IR attractive fixed point at $\tilde G=0$
and (if $a>0$) also an UV attractive nontrivial fixed point at $\tilde G_*=1/16\pi a$.
The solution of the RG equation (\ref{planckrg}) is
\be
\label{runningplanck}
M_{pl}^2(k)=M_{pl}^2(0)+a k^2\ .
\ee
We see that for $k\ll M_{pl}(0)$ the dimensionful $G$ is constant,
while the dimensionless $\tilde G$ scales as $k^2$. 
This is the regime that we are familiar with.
On the other hand for $k\gg M_{pl}(0)$ the dimensionful $G$ scales
as $k^{-2}$ and the dimensionless $\tilde G$ is constant.
This is the fixed point regime.

The aforementioned body of work extends this simple argument
in various directions and provides rather compelling, though for
the time being not yet conclusive, evidence for asymptotic safety.
In spite of this, there have been occasional claims that gravity 
cannot conform to this picture, because it would contradict other 
expected features of the theory at high energy.
For example, there is an increasingly popular attitude according to which
gravity should not be viewed as a fundamental force but rather
as an ``emergent'' phenomenon. Past attempts in this direction
went under the name of ``induced gravity'' \cite{adler}
or ``pregeometry'' \cite{akama}.
Recently there has been a revival of these ideas following
a derivation of the gravitational field equations
as thermodynamic properties \cite{emergent}.
The asymptotic safety approach, which seeks UV completeness 
of a theory of the metric, seems to be
flatly incompatible with this point of view.
Another popular expectation is that in quantum gravity one cannot
meaningfully talk of arbitrarily short distances.
Since an asymptotically safe theory is purportedly valid
``up to arbitrarily high energies'', it would seem to be
in contradiction with this expectation.

The aim of this note is to discuss the status of asymptotic safety 
in relation to these issues, and hopefully thereby to eliminate 
some possible sources of misunderstanding.
In section II we will show that if we consider a trajectory that is not
asymptotically safe but close to one that is,
we get a picture of asymptotic safety as an 
``emergent'' low energy property of gravity.
Far from being incompatible, asymptotic safety could thus prove to be
a powerful way of formulating the notion of emergence.
For this, it would be wrong {\it not} to treat the metric
degrees of freedom as quantum fields.
In section III we will discuss some scattering Gedankenexperiments
where asymptotic safety could manifest itself.
In section IV we then argue that in those experiments
that actually probe short distances, there is a sense
in which asymptotic safety implies a minimal length.
The argument is of a purely quantum nature 
and {has nothing to do with microscopic black holes}.
Section V contains some further discussion and a summary of our conclusions.


\section{Asymptotic Safety and emergence}


The notion of Asymptotic Safety, as described in the preceding section,
is an adaptation to particle physics of ideas that are rooted in
statistical mechanics and in particular in the theory of critical phenomena.
In condensed matter physics, one typically starts from a microscopic theory defined
on a lattice. Wilson's RG transformations provide a way of extracting
large scale properties of the system from the microscopic theory \cite{wilson}.
In particular, at a critical point, the ratio between
the correlation lenght and the lattice spacing goes to infinity.
Such a point describes the state of matter near a second order phase
transition, where fluctuations on all length scales are important
and the system becomes scale invariant.
The macroscopic properties of the system near such a point turn out to
be governed by just a few parameters (``critical exponents'')
that can be calculated in principle,
and a crucial property is ``universality'', namely the fact that these
parameters appear to be the same for several different system,
and therefore are to some degree independent of details of the microscopic theory.
Such large scale, thermodynamic properties can be said, in a definite sense, 
to be ``emergent'', 
and the theory allows us to make detailed quantitative predictions 
that have been successfully compared to experiment in many cases.

In the particle physics transcription of these ideas,
the notion of ``emergence'' seems to have been lost.
UV completeness is used as a criterion to select the right theory,
and for its sake one is ready to fine tune infinitely many parameters 
({\it i.e.} the irrelevant couplings;
in the case of the Gaussian FP this corresponds to setting all the
nonrenormalizable couplings exactly to zero).
This is precisely the logic that led to the formulation of the standard model,
so there is no denying its power and usefulness,
but it seems to be quite opposite to the Wilsonian one.

In fact this difference in perspective is not as profound as it may seem.
Its origin lies in the fact that in particle physics we usually know 
a theory at some scale and we are interested in extending it towards higher energies.
Since the theory is defined, at least formally, on a continuum,
UV problems always loom large.
In condensed matter physics, one usually has a reasonably good understanding
of the microscopic degrees of freedom and their interactions,
and one is interested in deriving the macroscopic properties.
The UV is never an issue, because of the atomic structure.

One can easily switch from one viewpoint to the other,
just by reversing the direction of the flow
\footnote{In statistical mechanics the RG flow always goes from the UV to the IR:
since many microscopic starting points can give the same macroscopic properties,
there is loss of information along the flow.
But if, as we discussed in the previous sections, one was able to follow the RG
of {\it all} the couplings of the theory, then one could run the flow in either
direction without loss of information.},
and considering not only trajectories that are {\it in} the
UV critical surface but also those that are {\it near} to it.
To see this, let us assume that there is a FP in theory space 
with finite dimensional UV critical surface.
We do not specify at this stage whether it is the Gaussian FP
or some other FP, so for the time being the discussion is very general.
Let us choose the initial point of
the flow more or less randomly in theory space.
Because the irrelevant directions are IR attractive,
{when flowing towards lower energies}
the theory will be driven very rapidly towards the UV critical surface.
If the initial point lies exactly in the IR critical surface
(which requires that $d$ coordinates be precisely tuned)
the flow will hit the FP in the IR.
We are interested in the situation when the starting point is not
{\it in} the IR critical surface, but close to it.
In this case, the flow will come close to the FP, where it will slow down.
Because the relevant directions are IR repulsive,
it will eventually restart following closely the UV critical surface.
The more accurate was the initial tuning of the relevant parameters,
and the longer the flow,
the closer the RG trajectory gets to the UV critical surface in the IR.
So, from the point of view of a low energy observer,
the flow will look {\it as if it originated from the FP}.
See fig.1.

The equation of the UV critical surface, expressing the values
of all the coordinates in theory space as functions
of $d$ undetermined parameters, and the speed of the flow
near the FP (the critical exponents) are predictions
of the theory that one can in principle test againts experiment.
These are then, in a well defined sense, ``emergent'' properties
of the theory at low energy.
The notion of ``universality'' corresponds to
independence of (some of) these predictions from the details of the microscopic theory, 
namely independence from the starting point of the flow in theory space.

%
\begin{figure}
\begin{center}
{\resizebox{0.6\columnwidth}{!}
{\includegraphics{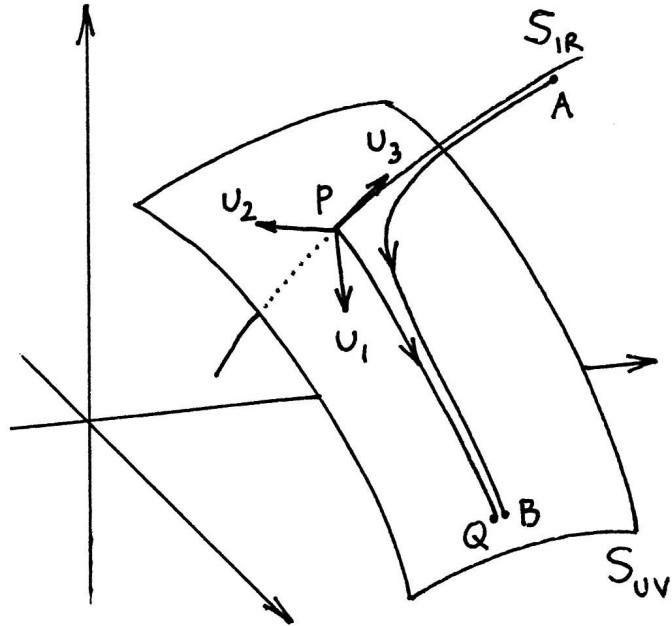}}
\caption{\small 
A view of theory space with a fixed point $P$,
with relevant directions $u_1$ and $u_2$, while $u_3$ is one of infinitely many
irrelevant directions.
$S_{UV}$ is the UV critical surface and $S_{IR}$ the IR critical surface.
Here $S_{IR}$ is drawn as if it was one-dimensional but in reality it is
infinite dimensional.
The trajectory $PQ$ is an asymptotically safe theory.
The trajectory $AB$ is an EFT.
Arrows run from high to low energy. The point $A$ corresponds to a
trans-Planckian value of $k$, the point $B$ to sub-Planckian $k$.
}
}
\end{center}
\end{figure}

At this point proponents of the view that gravity is emergent will feel 
uncomfortable with the fact that we have restricted
ourselves to the theory space of diffeomorphism invariant functionals of a metric.
They usually protest that it is conceptually wrong to start from the metric 
as a degree of freedom (d.o.f.), so they will see this choice 
as arbitrary and unmotivated.
Of course it is possible that at a more fundamental level
the microscopic d.o.f. are
not a metric, but it is also undeniably true that at low enough energies
gravity can be described by a metric, so at some point one has to make
a connection between the two descriptions.
This could work roughly as follows:
once the microscopic d.o.f. are specified,
one has to construct a metric out of them, run the RG and find that
at some scale the metric becomes a propagating field.
Then one performs a change of variable in the functional
integral in such a way that fluctuations over the metric are summed over.
The microscopic d.o.f. may decouple, or else remain in the form of matter fields.
From that point on, the theory will be described by an RG trajectory
in the theory space of metrics, or metrics and matter,
of the type that we have discussed above.
So from the point of view of low energy, all that the
microscopic theory adds to the preceding discussion is to provide 
the initial point of the RG flow.
This argument also shows that it would be wrong not to consider
the quantization of the metric d.o.f.,
even if in this scenario one would only regard it as an EFT.

Let us now distinguish the Gaussian from the non-Gaussian FP.
The eigenvalues of the linearized flow near the Gaussian FP are
given simply by the canonical mass dimension of each coupling.
The only relevant coupling is the cosmological constant. 
The couplings in the $R^2$ terms are marginal
and all the others are irrelevant.
If we want an RG trajectory that is asymptotically safe, we have to tune
all the couplings to lie in its UV critical surface.
In the immediate vicinity of the Gaussian FP,
this means putting all the irrelevant couplings to zero,
including Newton's constant.
But one thing that we know for sure about gravity
is that $\tilde G$, however small, is not zero.
So the real world cannot lie on such a trajectory.

The nonrenormalizability of the theory then manifests itself 
in the following way. Suppose we start near the Gaussian FP with
all couplings equal to zero except for a small $\tilde G$,
and we try to follow the trajectory for growing $k$.
Since $G$ is irrelevant, $\tilde G$ will grow and we will 
be pushed out of the domain of perturbation theory.
If there was no other FP, the trajectory would go
towards infinity, generating a divergence.
A small $\tilde G$ will also give rise to small but nonzero beta functions
also for the other irrelevant couplings.
These will then all grow and give rise to infinitely many divergences.
Clearly the theory stops making sense.
The energy scale at which $\tilde G$ becomes of order one is the
Planck scale. In this case the scale of new physics is the Planck scale
and the new, more fundamental d.o.f. must manifest themselves at this scale.
This is the classic picture of Einstein theory as an effective field theory
breaking down at the Planck scale.

On the other hand, if a nontrivial FP exists, as the arguments
in section I seem to suggest, the EFT can be used also beyond the Planck scale.
Since we know that $\tilde G$ is very small at low energy,
we must choose the RG trajectory so as to satisfy this condition.
One way in which this could happen is if the Gaussian FP
was contained in the UV critical surface of the nontrivial FP.
Then there would be a special trajectory that joins the
nontrivial FP in the UV to the Gaussian FP in the IR,
and the trajectory describing the real world would be close to this
trajectory for a very large range of scales.
{Such a trajectory exists
in the Einstein-Hilbert truncation} \cite{trajectory}
but it has so far been impossible to find it in more complete truncations,
possibly just for technical reasons.
It is important to realize that this aesthetically desirable
situation is not really required by existing observational data:
already for the $R^2$ terms there is essentially no significant bound,
much less for the higher ones.
So it is only required that $\tilde G$ becomes sufficiently small.

To summarize, the difference between the notion of asymptotic safety,
as presented in the preceding section,
and the standard Wilsonian point of view, with its associated notion of emergence,
is that in the former one initially tunes all the (infinitely many)
irrelevant couplings exactly to zero and runs the RG towards the UV,
whereas in the latter one initially tunes the (finitely many) 
relevant parameters and runs the RG towards the IR.
By putting the initial point near the IR critical surface,
the trajectory can be made arbitrarily close to the UV critical surface 
at some low energy scale.
In this way the theory is not strictly asymptotically safe,
but it is approximately so at low energy,
and this approximate asymptotic safety is an emergent property.
Insofar as, even in principle, no one will ever be able 
to reach infinitely high energies, nor to measure couplings
with infinite precision, the difference between the two 
points of view is physically immaterial.


\section{Scattering signatures of asymptotic safety}


There is a popular argument saying that a minimal length must arise 
in QFT when gravity is taken into account.
It is based on a variant of Heisenberg's microscope:
any attempt at measuring position with precision $\Delta x$
will put an energy of order $1/\Delta x$ in a region of volume $\Delta x^3$.
When $\Delta x$ is smaller than the Planck scale, it is also
smaller than the classical Schwarzschild radius associated to this energy,
so any attempt at producing a more accurate localization 
is expected to fail because a black hole is produced.
The most precise formulation of this argument has been given
in \cite{doplicher}, where its relation to spacetime uncertainty relations
and noncommutativity has also been discussed.
This intrinsic minimal length is then argued to provide a universal
cutoff rendering all QFTs finite.
This reasoning has been used recently to argue
against asymptotic safety \cite{dvali}.

The obvious weakness of these arguments is the assumption
of the validity of Einstein's equations at the Planck scale, 
whereas it is precisely modifications of Einstein's theory that
we expect to find in that regime. 
Thus, the formation of a black hole 
is not at all a foregone conclusion.
In particular, if gravity is asymptotically safe, it becomes weaker at 
short distances in such a way that a horizon may never form
\cite{falls,basu}
\footnote{roughly speaking, the Schwarzschild radius contains a factor $G$ 
which in the fixed point regime goes like $\Delta x^2$. 
Instead of increasing when $\Delta x$ decreases, the quantum-corrected 
Schwarzschild radius decreases like $\Delta x$.}.
The argument however depends on the value of coefficients of order one,
over which there is some uncertainty.
We will return to this point in the discussion.

There have also been claims in the literature that
hypothetical gravitational scattering experiments, typically in the ultra planckian regime,
would be well described either by the eikonal approximation or, 
in the strong coupling regime, by semiclassical black hole metrics.
In both cases the dominant contribution would be associated
to soft graviton exchange, preventing a direct test of asymptotic safety
\cite{giddings}.
We do not agree with this conclusion in general, because it is based on several
approximations in a framework which would be modified in the presence of
asymptotic safety.
While reaching definite conclusions would require much more detailed investigations, 
we will now discuss scattering processes,
constrained by kinematics and quantum numbers, 
that could allow us in principle to detect signs of asymptotic safety.
In the next section we will argue that in such experiments
one will also encounter a minimal length.

Let us start by considering as matter two
different flavours of scalar or fermion fields.
We consider a flat spacetime on which one prepares asymptotic
scattering states which are then allowed to interact
purely gravitationally in some specific kinematical regime.
We assume that the dynamics is described by some Wilsonian effective action
depending on a suitable momentum scale $k$. This effective action already
contains the effect of loops with momenta larger than $k$,
so in calculating the cross sections it can be used at tree level, provided
one is taking a completely general parametrization of the action in the theory space.
We further assume that there is a fixed point with the properties that are
required for asymptotic safety, and we begin by considering
the contribution of the Einstein-Hilbert term with a $k$-dependent Newton's constant.
We shall discuss below the extension to more complicated truncations.

By measuring the energy dependence of the cross sections,
and knowing how the $S$ matrix depends on the couplings in the effective action,
one can obtain information on the dependence of the couplings on
the typical scale of the particular interaction, in
our case the virtualities involved in the graviton interactions.

Both spin-0 and spin $\frac{1}{2}$ fields lead to similar conclusions. 
We consider in particular two kinds of processes:
(a) the scattering of two different flavored particles and 
(b) the annihilation of a particle-antiparticle of one flavor followed by the
production of a pair of a different flavor.

Consider first the tree level scattering of
two different massless particles induced by one graviton exchange. 
For scalars the squared invariant amplitude is:
\be
\label{scalarscatt}
|M_{0\,scatt}^{(0)}|^2=\frac{1}{4M_{pl}^4} \frac{s^2u^2}{t^2}\ .
\ee  
For fermions, summing over incoming spin states and averaging over
final spin states, it is:
\be
\label{spinorscatt}
|M_{\frac{1}{2}\,scatt}^{(0)}|^2=\frac{1}{128M_{pl}^4} 
\frac{s^4-6s^3u+18 s^2  u^2 -6su^3+u^4}{t^2}\ .
\ee  

From these expressions by crossing ($t \to s$, $u\to t$ and $s\to u$)
one immediately obtains the corresponding expressions for the annihilation
process:
\be
\label{scalarann}
|M_{0\,ann}^{(0)}|^2=\frac{1}{4M_{pl}^4} \frac{t^2u^2}{s^2}\ ,
\ee  
\be
|M_{\frac{1}{2}\,ann}^{(0)}|^2=\frac{1}{128M_{pl}^4} 
\frac{t^4-6t^3u+18 t^2 u^2 -6tu^3+u^4}{s^2}\ .
\ee
Having calculated these amplitudes at tree level with the scale-dependent 
effective action, one should now set the relevant scale for the running Planck mass. 
Similarly to collider processes such as deep inelastic scattering 
or $e^+e^-\to hadrons$,  in these processes with single graviton exchange one
should use $M_{pl}(t)$ for the scattering and $M_{pl}(s)$ for the annihilation processes.
Using $s=E_{cm}^2$, $t=-\frac{s}{2}(1-\cos{\theta})$ and 
$u=-\frac{s}{2}(1+\cos{\theta})$, one finds the differential cross section
\begin{equation}
\label{dcs}
\frac{d\sigma}{d\cos{\theta}}=\frac{1}{32\pi E_{cm}^2}|M|^2
\end{equation}  
as a function of the center of mass energy and scattering angle.
Let us consider for example the total cross section for the annihilation
process. After integrating the two (very different) 
angular distributions, one finds for scalars and fermions:
\be
\sigma_{0\,ann}^{(0)}=\frac{s}{1920\pi M_{pl}^4(s)} \quad , \quad
\sigma_{\frac{1}{2}\,ann}^{(0)}=\frac{s}{5120\pi M_{pl}^4(s)}\ .
\ee
We now use formula (\ref{runningplanck})
for the running Planck mass: $M_{pl}^2(s)=M_{pl0}^2+a \,s$.
Measuring all momenta and lengths 
in units of the ``low energy'' {\it non running} Planck mass $M_{pl0}$, 
we see that the cross section stops growing 
at a maximum for $E_{cm}= M_{pl0}/\sqrt{a}$ and thereafter decreases to zero as
$1/(a^2 s)$ when the $E_{cm}\to \infty$.
On the other hand, if we decide to measure all the quantites in units of the
running Planck mass at scale $E_{cm}$, 
which is the relevant dimensionful scale in this UV regime, 
both the center of mass energy 
and the cross section become constant, 
the latter settling to a fraction of the Planck area.
In Fig. 2 we compare the cross sections for fixed gravitational constant
and for the running case with two different values of $\tilde G$ at the
fixed point. 

\begin{figure}
\centering
\includegraphics[]{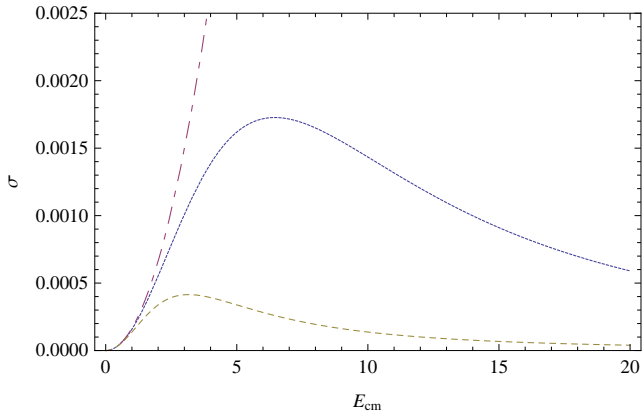}
\caption{Scalar annihilation cross section at tree level, in units 
of the low energy Planck mass $M_{pl0}$: the dot dashed line refers
to no asymptotic safety; the continuous and dashed lines
to asymptotic safety with a fixed point $\tilde M_{pl}^*=0.024$ and
$0.1$ respectively.}
\label{}
\end{figure}

The scattering amplitudes in (\ref{scalarscatt}) and (\ref{spinorscatt})
diverge for $\theta=0$ and the total cross sections also diverge.
If one cuts off the angular integration at some angle $\epsilon\to 0$,
the result is dominated by forward scattering.
In this case $t$ is small and one does not probe the asymptotic safety regime,
in agreement with \cite{giddings}.
However, if one considers the differential cross section at some fixed large angle,
both for spin 0 and spin 1/2,
its behavior as a function of $E_{cm}$ is very similar to the one
described above for the total annihilation cross section:
it grows quadratically if one treats $M_{pl}$ as a constant,
but if we set $M_{pl}^2(t)=M_{pl0}^2+a \,t$
it reaches a maximum for $E_{cm}\approx M_{pl0}$ and thereafter
decays to zero like $E_{cm}^{-2}$.
This is then the main conclusion of this discussion. 
Aside from the case of forward scattering,
asymptotic safety has a very dramatic effect:
it reverses the high energy behavior of the cross sections, 
which goes to zero like $E_{cm}^{-2}$ instead of growing like $E_{cm}^2$.

In this calculation we have neglected the contributions 
of the higher derivative terms and nonlocal terms.
The latter, which in a perturbative framework are the typical remnant 
of loop integrations and correspond to multigraviton exchanges,
could be important at high energy insofar as the theory becomes 
strongly interacting.
Such terms could be tackled in two different ways:
one way would be to retain them in the flow equation, along the lines of \cite{satz};
the other to first derive a bare action,
for example along the lines of \cite{manrique2}, and then use it to calculate loop effects.
To discuss the general case, let us parametrize the action as
\begin{equation}
\label{seff}
\Gamma_k=\sum_i g_i(k) \mathcal{O}_i(g_{\mu\nu},\psi)
\end{equation}
where $\mathcal{O}_i$ are integrals of functions of the metric $g_{\mu\nu}$,
the matter fields $\psi$ and derivatives.
The canonical dimension of the coupling $g_i$ is $d_i$,
and we define $\tilde g_i=g_i k^{-d_i}$.
The complete scattering amplitude appearing in (\ref{dcs})
must be constructed from this running effective action
taking into account the effect of all the terms in the action,
including the nonlocal ones.
It is a dimensionless function of the kinematical variables $s$, $t$,
of all the couplings $g_i$ and of $k$,
and therefore can be written as 
$|M|^2=f(s/k^2,t/k^2,\tilde g_i)$.
In a kinematical setup such as the one considered above,
which is characterized by a single scale $E_{cm}$,
we can set $k=E_{cm}$.
Then the differential cross section is
\begin{equation}
\frac{d\sigma}{d\cos\theta}=
\frac{f\left(\theta,\tilde g_i\right)}{E_{cm}^2} \ ,
\end{equation}
At fixed $\theta$, in the asymptotic safety regime the $\tilde g_i$
are constants and the asymptotic behavior of the cross section is  $E_{cm}^{-2}$,
exacly as in the preceding calculation.
Of course this argument cannot exclude that in some circumstances the function $f$ has some
nontrivial dependence on $E_{cm}$.
This is indeed what happens at low energies where the significant dimensionless couplings
$\tilde g_i$ run as $E_{cm}^{-d_i}$ 
({\it e.g.} $\tilde g_1=\tilde M_{pl}^2\sim E_{cm}^{-2}$) 
and the cross section goes, at tree level, like $E_{cm}^2$.
However, it is clear that the cross sections will generically differ
in the neighborhood of a fixed point, where the $\tilde g_i$ are constants,
from other regions of theory space where they depend on $E_{cm}$.
This provides in principle the possibility to establish experimentally
whether a theory is asymptotically safe.

The task of actually proving that gravity behaves in this way is of course
a very hard one. Assuming that Newton's constant has a fixed point,
and thinking of the behavior of the nonlocal,
multigraviton interactions, one could imagine that in a perturbative framework,
large contributions in the numerators of the Feynman integrals to be compensated, 
when virtualities reach the Planck scales, by large running Planck masses.
Moreover splitting of virtualities among many gravitons may be associated
to slower running, but also to smaller invariant energies in the
vertices, in such a way that corrections to tree level or lower loop order
diagrams appear to be important only at higher energies, essentially changing
the whole energy dependence in accordance to the asymptotic safety paradigm.
On the other hand dynamical features beyond the perturbative approximation
could be essential.
We leave all this for further future investigation.


\section{Asymptotic Safety and Minimal length}


As in the section 2, we begin within the framework of an EFT
with an UV cutoff $k$.
It is clear that even though the theory is formally described
on a continuum, as long as the cutoff is in place,
one cannot resolve distances shorter than $1/k$.
As is customary in lattice theory, we can take $k$ as unit of mass
and measure every dimensionful quantity in units of (powers of) $k$.
For example, Newton's constant in cutoff units is $\tilde G=G k^2$.
In section 1 we also gave a simple argument showing that $\tilde G$ has an UV FP.
Assuming that this is the true behavior of Newton's constant,
and that all other couplings in a theory of gravity behave similarly,
it would seem that one can take the limit $k\to\infty$, 
and hence that one could resolve arbitrarily short distance scales.
But is this statement physically meaningful?
The point one has to bear in mind is that a dimensionful quantity such as $k$
does not have any intrinsic value: it is only when we measure it
in some unit that we can attribute it a value.
So far we have used $k$ itself as a unit, but $k$ is always equal to one
in units of $k$, so in order to give a meaning to the statement ``$k\to\infty$''
we have to choose some other unit.
For example we can use Planck units, where the value of $k$ is $k\sqrt{G}$.
Since $G$ is a running coupling, one should specify at what scale
it is to be evaluated. 
This is a delicate issue whose answer depends on the physical
question that is being addressed.

If one wanted to measure the size of very small objects,
then arguably the $G$ that is relevant would not be the
one that is measured at low energy but rather the one that
would be measured {\it at the scale of the experiment}.
For example in a fixed angle scattering experiment of the type discussed in the
preceding section, the resolving power is determined by the
momentum transfer $\sqrt{t}$, 
and the natural unit to be used is $M_{pl}(k=\sqrt{t})$.
Borrowing terminology from \cite{reuter4}, 
we will refer to such measurements as ``proper''.
The ``proper'' value of the cutoff in Planck units is $k\sqrt{G(k)}$.
But $k\sqrt{G(k)}=\sqrt{\tilde G(k)}$, so it is essentially a tautology that in an asymptotically safe theory $k$ has an upper bound when 
``properly'' measured in Planck units \cite{perini3,percacci}.
It helps to think of the way in which this bound would be attained.
At all presently attainable energies, $\tilde G$ is a very small number.
In this regime $G$ can be thought of as being constant and
$k\sqrt{G}$ grows linearly with $k$.
Near the Planck scale $\tilde G$ becomes of order one,
$G$ begins to run like $k^{-2}$ and $k\sqrt{G}$ becomes constant.
Since $k$ itself is the upper bound for the momenta that
one can talk of, one concludes that one cannot talk at all
of ``proper'' momenta greater than Planck mass,
or ``proper'' distances shorter than the Planck length.
It is important to realize that this does not prevent us from talking
of asymptotic scattering states with trans-Planckian energies.
Such particles are on shell and the only scale characterizing them
is the invariant mass.
In this case even a ``proper'' measurement can yield unbounded energy
(though not an unbounded resolving power).
Whenever we talk of trans-Planckian energies we mean energies of free particles
measured in this way, or otherwise measured in low-energy units.

Is this a peculiar issue with Planck units, or the symptom of a more general phenomenon?
Could one not choose another unit that did not behave this way? 
We could take any dimensionful coupling $g_i$ of (\ref{seff})
and choose $(g_i)^{\frac{1}{d_i}}$ as unit of mass.
But if the theory is asymptotically safe,
$g_i k^{-d_i}$ will tend to a constant in the UV, so
the same argument that we used for Planck units will go through here as well.
We see that in an asymptotically safe theory of gravity,
the cutoff $k$ has an upper bound in any possible system of units.

Reuter and Schwindt \cite{reuter4}
arrived at a similar conclusion by a slightly different route. 
They distinguish momenta/distances measured with respect to a fixed metric
from momenta/distances measured with a ``proper'' metric, 
{\it i.e.} a metric that solves the quantum equations of motion with the
couplings evaluated at the scale $k$.
When the scale reaches the Planck mass, the couplings begin to run strongly
with $k$, and so does the ``proper'' metric.
In particular, in the Einstein-Hilbert truncation, 
since the equations of pure gravity do not depend on $G$,
the ``proper'' metric is proportional to $1/\Lambda(k)$.
In the fixed point regime it runs like $k^{-2}$,
so when one measures an object of size $k^{-1}$ with this metric,
the result is independent of $k$.
In this approach the scaling of the ``proper'' metric has the
same effect as the scaling of the unit in the discussion above.

The relation between these two approaches becomes evident 
if we choose to work with dimensionless coordinates.
Then, the covariant metric has dimension 
of length squared, as required by the basic formula
$ds^2=g_{\mu\nu}dx^\mu dx^\nu$.
The metric must be of the form
$g_{\mu\nu}=\ell^2 \bar g_{\mu\nu}$ where $\bar g_{\mu\nu}$
is dimensionless and $\ell$ is a unit of length.
For example, in flat space we can choose $\bar g_{\mu\nu}=\eta_{\mu\nu}$.
This emphasizes that the definition of the metric contains implicitly a choice of units.

When one works with a dimensionful metric, raising and lowering indices changes the
dimensions of tensor components.
The canonical dimensions of fields are generally different from
the standard ones.
The kinetic term for a tensor field of type $(r,s)$
($r$ covariant and $s$ contravariant indices)
will contain terms of the type
\begin{equation}
\label{candim}\int d^4x \sqrt{-g}g^{\mu_1\nu_1}\ldots g^{\mu_r\nu_r}
g_{\rho_1\sigma_1}\ldots g_{\rho_s\sigma_s}g^{\alpha\beta}
\nabla_\alpha t_{\mu_1\ldots\mu_r}^{\rho_1\ldots\rho_s}
\nabla_\beta t_{\nu_1\ldots\nu_r}^{\sigma_1\ldots\sigma_s}\ .
\end{equation}
The dimension of $t_{\mu_1\ldots\mu_r}^{\rho_1\ldots\rho_s}$ is $1-r+s$.
This agrees with the standard dimension only for scalars.
For example, gauge fields are dimensionless, in accordance with the
fact that they appear in covariant derivatives on the same footing
with ordinary derivatives.
It is now easy to see that the canonical dimension of all couplings
are not modified relative to the standard case: the Yang-Mills coupling is
dimensionless, Newton's constant has dimension of area etc.

Now consider Planck units, where $\ell=\sqrt{G}$.
We expand $g_{\mu\nu}=G\eta_{\mu\nu}+h_{\mu\nu}$, 
so that the graviton field $h_{\mu\nu}$ is not canonically normalized:
its propagator in de Donder gauge is $G P^{(2)\mu\nu\rho\sigma}/q^2$,
where $P^{(2)}$ is the spin-two projector and $q^2=g^{\mu\nu}q_\mu q_\nu$
(recall that $q_\mu\sim i\partial_\mu$ is now dimensionless).
Now consider one of the fixed angle scattering experiments discussed in the previous section. In naive perturbation theory all metrics appearing in the Feynman rules
should be identified with the ``low energy metric'' $G_0\eta_{\mu\nu}$
(in particular $p^2=G_0^{-1}\eta^{\mu\nu}p_\mu p_\nu$),
but we have seen that when we evaluate the scattering amplitude, 
the overall factor of $G$ should be evaluated at the scale $k=|q|$.
If we trace the origin of that factor, we see that it comes from the graviton propagator,
which we can write alternatively as
\begin{equation}
\frac{G(|q|)P^{(2)\mu\nu\rho\sigma}}{G_0^{-1}\eta^{\mu\nu}q_\mu q_\nu}
=\frac{G_0 P^{(2)\mu\nu\rho\sigma}}{G(|q|)^{-1}\eta^{\mu\nu}q_\mu q_\nu}\ .
\end{equation}
Note that since the dimensionless momenta $p_\mu$ of the external particles 
are measured in ``low-energy'' Planck units, the metric appearing in the
spin-two projector, which is used to contract them,
is the ``low energy'' metric $G_0^{-1}\eta^{\mu\nu}$.
We can thus interpret the RG improvement by saying that
the virtual graviton propagates in a scale-dependent background metric
\begin{equation}
\label{propermetric}
g_{\mu\nu}(|q|)=G(|q|)\eta_{\mu\nu}\ .
\end{equation}
It is this metric that is relevant when probing small objects,
and not the macroscopic metric $g_{\mu\nu}(0)=G_0\eta_{\mu\nu}$.
In the fixed point regime, this metric runs like $q^{-2}$,
exactly as discussed by Reuter and Schwindt, with the result that any
structure which would have sub-Planckian size when measured in units of $G_0$,
is magnified and appears to have Planck size with respect to the ``proper'' metric.
The ``proper'' resolving power of such an experiment is thus given by 
(a multiple of) the Planck length.
It is interesting to note that the running metric can give rise 
to deformations of the action of the Lorentz group \cite{hossenfelder,girelli,chp}.

Finally, we ask whether this conclusion could be avoided by using
a unit that is not derived from the gravitational sector, 
for example SI units.
We note that such units are based on atomic spectroscopy,
where the scale is determined by the mass of the electron,
which in turn is derived from the Higgs VEV.
Since here we set $\hbar=c=1$, for our purposes this is equivalent
to chosing the mass unit as a fixed multiple of the Higgs VEV.
It is clear that such units cannot be useful up to arbitrarily
high energies. For example, it could happen that at high temperature
the theory undergoes a transition to a phase where the VEV is zero.
Or, more to the point, if the theory is asymptotically safe, 
beyond a certain threshold
the VEV will begin to scale linearly with the cutoff \cite{giessm}.
Then we would find that 1/TeV could act as a minimal length 
in such ``electroweak'' units.
This would not
signal an absolute minimal length, because one could still use 
the gravitational couplings to measure shorter distances.
But we do not know of any mass scale higher than the Planck
scale, so in that case the breakdown of the ``gravitational'' units
would actually imply an absolute limit.

This argument for the existence of a minimal length
is very strange for someone who is accustomed to Einstein's Relativity.
In that theory it is assumed that it is always possible to make measurements
of distances and time intervals by using suitable rods and clocks.
But the properties of these objects are not explained within the theory itself. 
In particular the properties
of the macroscopic rods and clocks that Einstein had in mind are governed
mainly by electromagnetic interactions (for a recent discussion see \cite{brown}).
Here we are discussing a fundamental theory, which should make sense
by itself, without reference to external notions.
Especially if this theory also contains the strong and electroweak forces,
the properties of the rods and clocks, and the units they carry,
have to be defined entirely within the theory.
It is this requirement, together with the restriction to ``proper'' 
measurements, that leads to a minimal length.


\section{Discussion}


The very definition of asymptotic safety of gravity, with its emphasis on UV completeness,
and the RG calculations performed using metods of continuum, covariant QFT,
seem to lead to an interpretation of the theory as
(1) a fundamental QFT of a metric 
(2) defined on a physical continuum.
We hope to have convinced the reader that statement (1) is just one of two
possible interpretations.
By considering a trajectory that is nearly, but not exactly, asymptotically safe,
one would describe an EFT whose low energy properties, are ``emergent'', 
in the sense that they are nearly independent of the actual UV completion, 
and can be calculated within the EFT itself.
Such an EFT would be almost as predictive as an asymptotically safe one,
in the sense that the couplings are expected to lie {\it near} the UV critical surface
at and below the Planck scale.
The UV completion may well be in terms of non-metric d.o.f..
The endpoint of the flow at low energy is to a large extent independent of the initial point, 
so the nature and properties of the microscopic d.o.f. will remain unfathomable
to sub-Planckian observers.
At best one could hope, by precisely measuring the displacement of the actual trajectory
from the UV critical surface, to deduce the energy scale at which the
trajectory departs from asymptotic safety in the UV
\footnote{This is a generalization of the familiar argument that allows us,
in an EFT near the Gaussian FP, to calculate the scale of ``new physics''
from the coefficient of nonrenormalizable operators.}.
If the Gaussian FP is the only one, then this scale is the Planck scale.
If there is another FP, it could be much higher than that.

Let us now examine statement (2) from the ``fundamental'' point of view,
{\it i.e.} for an asymptotically safe trajectory.
Independent of any arguments of the ``Heisenberg microscope'' type,
the very definition of asymptotic safety implies that 
if one restricts oneself to ``proper'' measurements, one cannot probe distances
shorter that the Planck scale.
The reason is very simple: since the theory is ``fundamental'', unlike the
analogous situation of critical phenomena in condensed matter physics,
we cannot appeal to any external unit of mass or length.
The unit has to be chosen within the theory, and in the FP regime all the
possible candidates appear in constant, finite ratios
between themselves and the cutoff.
In this sense one could never probe a truly ``trans-Planckian'' regime.
This conclusion may seem a bit less perplexing if one recalls
that at the FP one has scale invariance, and that in a fundamental, 
scale invariant theory one cannot talk of distances.
One can only meaningfully speak of distances in the low energy,
subplanckian regime, and in that regime the shortest distance is
the Planck distance.

Now let us consider statement (2) from the point of view of ``emergence'',
{\it i.e.} for a trajectory that is close to asymptotic safety but not exactly safe.
The closer such trajectory gets to the FP, the longer the time it passes there.
Going upwards with energies, we assume that the scale at which it comes close
to the FP is $k=m_P=1/\sqrt{G}$,
and the scale at which it departs again from the FP is the scale of new physics $k=m_{NP}$.
This is the scale where new, presently unknown, pregeometic d.o.f.
and new interactions would manifest themselves.
The RG flow of $m_{NP}$ is expected to be suppressed for $k<m_{NP}$,
so in this regime one can use $m_{NP}$ as a reliable unit and
use it to make sense of sub-Planckian distances.
But then, this being an EFT, we have a new minimal length $1/m_{NP}$.
So, if we regard asymptotic safety as an approximate low energy property
of an EFT, and if the Heisenberg microscope argument does not apply,
one can actually speak of sub-Planckian distances,
down to the characteristic length scale of the new interactions.

This ``Wilsonian'' picture of gravity has been recently called into question 
based on the assumption that in the trans-Planckian regime black hole formation will 
dominate the dynamics (see \cite{giddings,dvali} and references therein).
We find this kind of reasoning unconvincing, for several reasons.
First of all, even when one follows the logic described in those papers,
it is not at all obvious that when one tries to resolve
sub-Planckian structures a black hole will form.
If one is willing to restrict oneself to one graviton exchange, as in~\cite{dvali}, 
it is clear that the gravitational coupling should be evaluated
at the energy scales involved in the process.
Then, the Schwarzschild radius
that enters in the discussion of the Heisenberg microscope (see sect. III)
scales linearly with $\Delta x$ when $\Delta x$ is sufficiently small,
and like $(\Delta x)^{-1}$ when $\Delta x$ is large.
The formation of a black hole depends on the details of the behavior of the Schwarzschild radius for $\Delta x$ comparable to the Planck length,
which is itself of the order of the Planck length.
Second, we should remember that black holes, even of the astrophysical variety,
are still speculative objects. We do not feel that their conjectured properties,
especially when extrapolated to the microscopic domain,
are sufficiently well established to draw firm conclusions.
And third, even if the classical argument for the formation of a black hole
was correct, its significance in the quantum context is far from obvious.
Before solving the field equations one should make sure that they do not
receive significant quantum corrections, and even then, given that
quantum fluctuations of the metric are expected,
the behavior of individual classical solutions of the equations of motion is not necessarily important.

In contrast to the view of \cite{dvali} that the microscopic behavior
the theory is dominated by classical configurations,
asymptotic safety would represent a backreaction of the {\it quantum}
theory limiting the unbounded growth of fluctuations and averting the impending ultraviolet catastrophe.
The authors of \cite{dvali} also view asymptotic safety essentially as
a modification of the propagator and argue, always based on black hole formation,
that there cannot be any propagating states with poles at masses higher than the Planck mass.
This certainly excludes the scenario of gravity as an emergent phenomenon, 
but not asymptotic safety.
On the contrary, it is amusing to observe that asymptotic safety 
(of the ``fundamental'' variety) also predicts
that there cannot be poles in propagators at masses
greatly exceeding the Planck mass. The reason is that for such a candidate
trans-Planckian mass state, the running mass will grow like $p^2$ 
with some coefficient greater than one, and hence can never be equal to $p^2$.
This too can be seen as a consequence of scale invariance.

A more difficult issue is the one raised in \cite{giddings}, 
namely the calculation of multigraviton exchange, possibly at large couplings.
This suggests including in the truncation specific non local terms, 
also involving matter, and studying their RG flow.
This will require a major effort.
In any case, the arguments given in section III show that 
in an asymptotically safe theory the cross sections 
will have an asymptotic $E_{cm}^{-2}$ behavior, 
which in principle provides a characteristic signature.

If the world is described by an asymptotically safe QFT,
or one that is close to asymptotic safety,
then not only the gravitational couplings but also the couplings 
of the standard model (or perhaps a GUT) must approach the FP.
It is natural to expect that this happens in the order 
set by the characteristic scales of the three interactions.
The strong coupling slowly moves towards asymptotic freedom
starting at scales of order of one GeV.
One would expect the electroweak force to follow suit
at the TeV scale and the gravitational couplings to
reach the FP at the Planck scale.
In an asymptotically safe standard model the Higgs VEV is expected to
become conformal at the Fermi scale \cite{giessm}.
Perhaps the most exciting possibility would be that the onset of the conformal regime 
manifests itself at the TeV scale in some form that is detectable at the LHC,
for example a suppression of the Higgs resonance.

\medskip
\centerline{Acknowledgements}
RP wishes to thank the Perimeter Institute for hospitality in the early stages
of this work, and D. Benedetti, S. Giddings, R. Gurau, S. Hossenfelder,
T. Padmanabhan, L. Smolin, E. Verlinde for discussions.

\noindent

\goodbreak

\end{document}